\documentclass[aps,prl,twocolumn,showpacs,superscriptaddress,preprintnumbers,amsmath,amssymb]{revtex4}

\usepackage{graphicx}
\usepackage{dcolumn}
\usepackage{bm}
\usepackage{color}
\usepackage{multirow}
\usepackage{textcomp}

\begin{document}


\title{Relevance of Abelian Symmetry and Stochasticity in Directed Sandpiles}
\author{Hang-Hyun Jo}
\affiliation{School of Physics, Korea
Institute for Advanced Study, Seoul 130-722, Korea}

\author{Meesoon Ha}
\email[Corresponding author: ]{msha@kaist.ac.kr}
\affiliation{Department of Physics, Korea
Advanced Institute of Science and Technology, Daejeon 305-701,
Korea}

\date{\today}

\begin{abstract}
We provide a comprehensive view on the role of Abelian symmetry
and stochasticity in the universality class of directed sandpile
models, in context of the underlying spatial correlations of
metastable patterns and scars. It is argued that the relevance of
Abelian symmetry may depend on whether the dynamic rule is
stochastic or deterministic, by means of the interaction of
metastable patterns and avalanche flow. Based on the new scaling
relations, we conjecture critical exponents for avalanche, which
is confirmed reasonably well in large-scale numerical simulations.


\end{abstract}

\pacs{05.65.+b, 05.70.Ln, 64.60.Ht}



\maketitle

Since a prototype of sandpile models was first introduced by Bak,
Tang, and Wiesenfeld, lots of its variants have been tested and
become successful over the last two decades in figuring out the
underlying common mechanism of ubiquitous scale invariance in
nature~\cite{BJD}. In such models, grains are slowly added,
redistributed (toppled) instantly whenever the instability
threshold is overcome, and finally dissipated at boundaries. It
has been the most interesting issue and under debate whether the
universality class of critical avalanche dynamics can be changed
by the modification of local toppling rules, such as the breaking
of Abelian symmetry~\cite{YCZhang1989} and the consideration of
stochasticity~\cite{Manna1991}, with still conflicting numerical
results~\cite{B-HLLMD}. Abelian symmetry here means that the order
of toppling the unstable sites does not affect the final state.
Contrary to undirected models, directed sandpile models (DSMs)
with a preferred direction of toppling turns out to be more
tractable analytically as long as they have Abelian
symmetry~\cite{Dhar1989,Pastor-Satorras2000,PK}. It is because
metastable patterns in the Abelian DSMs are fully uncorrelated.

Once the Abelian symmetry is broken in DSMs by some specific way,
long-range spatial correlations emerge in their metastable
patterns. Such correlations are often observed in nature, like a
fractal structure in the crust of the earth formed by seismic
events. Two non-Abelian DSMs with spatially correlated metastable
patterns were introduced by Hughes and Paczuski~\cite{Hughes2002}
for a stochastic version and by Pan~\textit{et al.}~\cite{Pan2005}
for a deterministic version. In the stochastic version it is
claimed that Abelian symmetry is not relevant to avalanche
dynamics, while in the deterministic version it is. Although this
difference might be attributed to the existence of stochasticity,
it is not clear enough to say which factor governs the scaling
property of metastable patterns. Therefore, it is quite crucial to
clarify the role of spatially correlated metastable patterns in
the universality class of DSMs, which has been hardly discussed up
to now.


In this Letter, we discuss how the critical avalanche dynamics of
non-Abelian models are entangled with spatially correlated
metastable patterns. Based on the formation of metastable patterns
and scars (trace of avalanche boundary sites) with the mapping
onto particle dynamics, we give intuitive arguments about the
scaling relations in terms of scar exponent, and conjecture a
possible scenario for the universality class in DSMs. Finally, we
reinterpret the earlier known results for the Abelian case by our
conjecture, and confirm those for non-Abelian case by large-scale
numerical simulations with various data analysis techniques
developed so far.

Consider DSMs defined on a ($1+1$)-dimensional tilted square
lattice of size $(L,T)$. The preferred direction of avalanche
propagation is denoted by the `layer' $t=0,\cdots,T-1$ with open
boundary conditions, and the transverse direction by
$i=0,\cdots,L-1$ with periodic boundary conditions. Initially, to
each site of the lattice an integer value (the number of grains),
$z_i(t)\in[0,z_c)$, is assigned, where we set the instability
threshold $z_c=2$. Given a stable configuration where all sites
are stable, new grains are added one by one at a randomly chosen
site on the top layer, $z_i(0)\rightarrow z_i(0)+1$, until one of
them becomes unstable. For any unstable site with $z_i(t)\geq 2$,
grains at that site topple down to its left and right
nearest-neighboring sites on the next layer, $t+1$:
\begin{eqnarray}
z_i(t)\to z_i(t)-\Delta_{ii},\nonumber\\
z_{i\pm 1}(t+1)\to z_{i\pm 1}(t+1)+\Delta_{i,i\pm 1},
\end{eqnarray}
where $\Delta_{ii}=\Delta_{i,i-1}+\Delta_{i,i+1}$ (the local
conservation of grains). Toppled grains at the unstable sites on
the bottom layer $t=T-1$ are dissipated out of the system. Only
after another stable configuration is recovered by a series of
toppling events, denoting an avalanche, a new grain is added to
keep generating another avalanche.
\begin{figure}[t]
\includegraphics[width=0.925\columnwidth,angle=0]{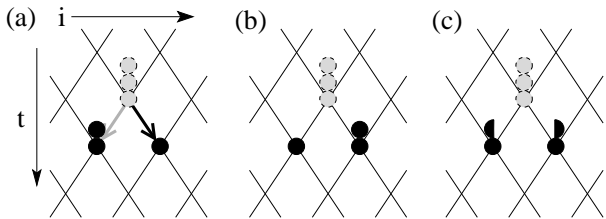}
\caption{Toppling rules of non-Abelian deterministic DSMs on a
($1+1$)-dimensional tilted lattice: (a)~aND, (b)~bND, and (c)~cND.
The gray colored grains and arrow represent the state before
toppling and the black colored ones represent the state after
toppling, respectively.} \label{fig:toppling}
\end{figure}
By setting $\{\Delta_{ij}\}$ one may consider several variants of
DSMs. In contrast to Abelian DSMs where $\Delta_{ii}$ is constant
(used to be set as $z_c$), we set $\Delta_{ii}=z_i(t)$ as
non-Abelian DSMs. All grains at the unstable site topple to the
next layer and the toppled site becomes completely empty. For any
given $\Delta_{ii}$, the values of $\Delta_{i,i\pm 1}$ can be
determined in either stochastic or deterministic way. Besides the
well-known Abelian deterministic or stochastic
DSMs~\cite{Dhar1989,Pastor-Satorras2000} (AD/AS in short) and the
non-Abelian stochastic DSM~\cite{Hughes2002} (NS in short), we
explore the following three versions as the non-Abelian
deterministic DSM (ND in short):
\begin{eqnarray}
{\rm (i)}~\Delta_{i,i\pm 1} &=&\left\{\begin{array}{ll}k & \textrm{if $z_i(t)=2k$}\\
k+\delta_{i\pm 1, a_i(t)} & \textrm{if $z_i(t)=2k+1$},\end{array}\right. \nonumber\\
{\rm (ii)}~\Delta_{i,i\pm 1}&=&\left\{\begin{array}{ll}k & \textrm{if $z_i(t)=2k$}\\
k+\delta_{i\pm 1, i+1} & \textrm{if $z_i(t)=2k+1$},\end{array}\right. \nonumber\\
{\rm (iii)}~\Delta_{i,i\pm 1} &=&z_i(t)/2,\nonumber
\end{eqnarray}
where $k$ is a positive integer and $\delta_{ij}$ denotes the
Kronecker delta-function. We call (i) the alternatively biased
version (aND)~\cite{Pan2005}, (ii) the fully biased version (bND),
and (iii) the continuous version without bias (cND), respectively.
For the aND, an `arrow' $a_i(t)$ of each site initially points to
one of its neighbors, say $i-1$ in Fig.~\ref{fig:toppling}~(a).
Whenever each grain is toppled at that site, the direction of the
arrow flips to the other neighbor, see Fig.~\ref{fig:toppling},
which shows the case of $z_i(t)=3$.

Each avalanche can be characterized by the following quantities:
mass $s$ (the number of toppled grains), duration $t$ (the number
of affected layers), area $a$ (the number of distinct toppled
sites), width $w$ (the mean distance between left and right
boundaries of avalanche), and height $h$ (the mean number of
toppled grains per toppled site). The avalanche distribution
functions in DSMs show no characteristic scale except for $T$ as
long as $L$ is sufficiently larger than the maximum width. They
follow the simple scaling form as $P(x)\sim x^{-\tau_x} f\left(
x/T^{D_x} \right)$ for $x \in \{s,t,a,w,h\}$. Moreover, two
quantities $x$ and $y$ scale as $\langle y\rangle \sim
x^{\gamma_{yx}}$ with $\gamma_{yx}=\frac{\tau_x-1}
{\tau_y-1}=\frac{D_y}{D_x}$ from $P(x)dx=P(y)dy$. Taking full
advantage of the relations, $D_t=1$ and $\langle s\rangle\sim T$
in DSMs, with the reasonable assumption of compactness of
avalanche, i.e. $a\sim wt$ and $s\sim ah$, we obtain the following
scaling relations: $\gamma_{xt}=D_x$ for any $x$,
$D_s(2-\tau_s)=1$, $D_a=D_w+1$, and $D_s=D_a+D_h$. As a result,
there are only two independent exponents left. Concerning the
metastable state, we define two scaling exponents more. Along the
propagation direction, one can measure grain density as
$\rho(t)\equiv \langle \frac{1}{L} \sum_i z_i(t)\rangle\sim
t^{-\alpha}$ with the grain density exponent $\alpha$ for large
$t$. The scar density $\rho_{\rm{sc}}(t)$ and the scar exponent
$\alpha_{\rm{sc}}$ are defined by the same definition of grain
density only with $z_i(t)$ replaced by $b_i(t)$, which takes a
value of $1$ for trace or $0$ otherwise. The scar exponent is
immediately related to the avalanche width exponent as
$\alpha_{\rm{sc}}=D_w$ because the density of avalanche boundary
sites is inversely proportional to the typical avalanche width,
i.e. $\rho_{\rm{sc}}(t)\simeq w(t)^{-1}$. Since grains can remain
only at the avalanche boundary sites for the non-Abelian DSMs, it
is found that $\alpha=\alpha_{\rm sc}$.

We give intuitive arguments on the interplay between avalanche
flow and metastable patterns/scars in DSMs. Let us define $N(t)$
as the number of grains transferred from the layer $t$ to the next
layer $t+1$ within an avalanche, scaling as $N(t)\sim w(t)h(t)\sim
t^{D_w+D_h}$. The evolution of $N(t)$, avalanche flow, can be
written as
\begin{equation}
\frac{dN(t)}{dt} \approx N(t)-N(t-1)=\sum_{i\in w(t)} n_i(t),
\label{eq:floweq}
\end{equation}
where $n_i(t)$ denotes the amount of the avalanche flow
at each site $(i,t)$, and the summation is over
the sites between avalanche boundaries belonging to $w$.

We begin with the Abelian case, for which the metastable patterns
are fully uncorrelated. In the AD, it is well-known that $D_h=0$
by definition and $D_w=1/2$ by mapping avalanche boundaries onto
the random walks~\cite{Dhar1989}. The avalanche flow of the AD can
be written as $\frac{dN}{dt}\approx \eta$. An uncorrelated noise
$\eta$ of zero mean and unit variance denotes the fluctuation of
grain density. In the AS with the same $\eta$, the bulk
contribution to avalanche flow plays a crucial role, so that we
get $\frac{dN}{dt}\approx
\sqrt{w}\eta$~\cite{Pastor-Satorras2000,PK}. Here $\sqrt{w}$
represents the fluctuation of the number of toppled sites in $w$
when the topplings are uncorrelated. The lack of correlation in
metastable patterns again leads to $D_w=1/2$, so
$D_h=\frac{1-D_w}{2}=1/4$. In relation to the scar density, we get
$\alpha_{\rm sc}=D_w=1/2$ but $\alpha=0$ for the Abelian case. It
is found that the measurement of $\alpha_{\rm sc}$ is more
efficient than that of $D_w$. From now on, we suggest to measure
$\alpha_{\rm sc}$ as one independent exponent in DSMs.

For the non-Abelian case with the spatial correlation of
metastable patterns or scars, the fluctuation of grain density is
numerically found to scale as $\rho(t)$. Furthermore, $\rho_{\rm
sc}$ plays the same role as $\rho$, i.e. $\alpha_{\rm sc}=\alpha$.
In the NS~\cite{Hughes2002}, keeping $\sqrt{w}$ due to the
stochastic nature of toppling just as in the AS, we can write
$\frac{dN}{dt}\approx \sqrt{w}\rho$ and get the new scaling
relation, $D_h=1-\frac{3}{2}\alpha$. In contrast to the NS, in the
ND, all sites of $w$ contribute to avalanche flow so that
$\frac{dN}{dt}\approx w\rho$. We get the different relation
$D_h=1-\alpha$. Therefore, all avalanche exponents in the
non-Abelian DSMs can be obtained from the grain density exponent
$\alpha$ (or $\alpha_{\rm sc}$) as shown in
Table~\ref{table-exponent}.

\begin{figure}[t]
\includegraphics[width=0.925\columnwidth,angle=0]{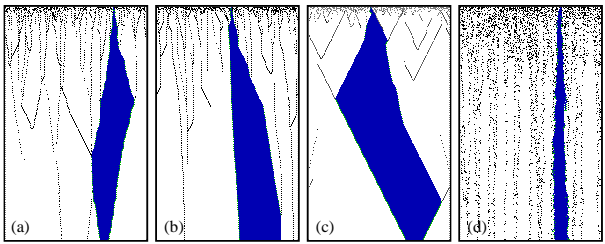}
\caption{(color online) Typical metastable patterns for
non-Abelian case: The occupied sites are shown as black dots and
the typical shapes of dissipative avalanches consisting of the
toppled sites as blue (or gray)-shaded areas on a lattice with $L=150$ and
$T=250$. Here (a) aND, (b) bND, (c) cND, and (d) NS,
respectively.} \label{fig:metastable}
\end{figure}
The new scaling relations with $\alpha$ enable us to clarify the
effect of metastable pattern with $\alpha\approx 0.45$ in the NS
on its avalanche dynamics. Interestingly, if $\alpha=1/2$ is
assumed with systemic errors and/or possible logarithmic
corrections, the NS has exactly the same avalanche exponents as
those of AS: $\tau_s=\frac{2(3-\alpha)}{(4-\alpha)}=10/7$ and
$\tau_t=2-\frac{\alpha}{2}=7/4$. Another scenario for $\alpha=1/2$
can be found by mapping metastable patterns onto the space-time
configuration of $2A\to A$ coagulation-diffusion model defined in
$d=1$, where the particle density decays as
$t^{-1/2}$~\cite{Hinrichsen2000}. One can say that the NS belongs
to the same universality class as the AS in the following sense:
For Abelian cases the flowing avalanche can sweep and lose many
grains at the same time due to the uniform grain density. On the
other hand, for non-Abelian case the flowing avalanche can sweep
only a few grains due to the power-law decaying grain density and
leave few grains behind by taking all grains at the toppling
sites. In other words, the scaling property of $N(t)$ is
apparently unaffected by the grain density as long as the toppling
rule is stochastic.
\begin{figure}[!b]
\includegraphics[width=0.725\columnwidth,angle=0]{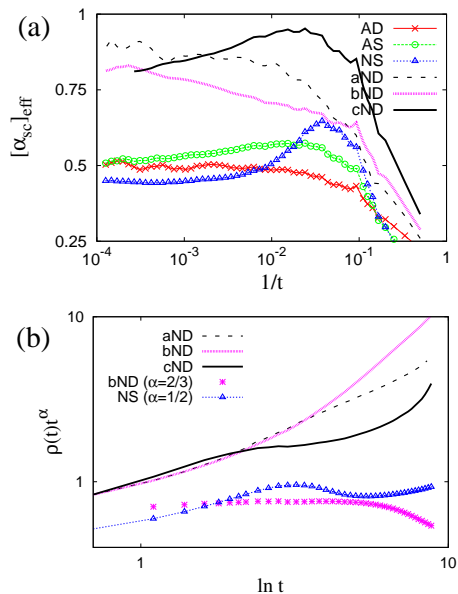}
\caption{(color online) (a) The effective scar exponent
$[\alpha_{\rm sc}]_{\rm eff}$ as a function of $1/t$, and (b)
double-logarithmic plots of $\rho(t)t^{\alpha}$ versus $\ln t$ for
logarithmic correction checks, where $\alpha=1$ unless noted. For
the bND, we cannot exclude the possibility of $\alpha=2/3$ either.
Here the data of each model were obtained from $10^8$ or more
avalanches on a lattice of $L=2^{14}$ and $T=L$, except for the
cND where $T=2^{13}$.} \label{fig:alphanrho}
\end{figure}
\begin{table*}
\caption{Avalanche exponents $\{\tau_x,D_x\}$, grain density
exponent $\alpha$ and scar exponent $\alpha_{\rm sc}$ in
($1+1$)-dimensional DSMs with our conjecture. Note that
$\alpha_{\rm sc}=\alpha$ for the non-Abelian case. We check the
value of $K_x\equiv D_x(\tau_x-1)$, which is a universal constant
for any $x$~\cite{Lubeck2000}, and show its averaged value over
$x$ excluding $K_w$ and $K_h$ due to their poor statistics.}
\label{table-exponent}
\begin{tabular}{l*{8}{c}}
\hline\hline Model & $\tau_s$, $D_s$ & $\tau_t$, $D_t$ & $\tau_a$,
$D_a$ & $\tau_w$, $D_w$ & $\tau_h$, $D_h$ & $\langle
K_x\rangle$ & $\alpha$, $\alpha_{\rm sc}$ \\
\\

Mean field ($d_u=2$) & $3/2$,~$2$ & $2$,~$1$ & $3/2$,~$2$ &
$3$,~$1/2$ &
$\infty$,~$0$ & $1$ &\\
\\

Abelian\\
$~~$
Deterministic & $4/3$,~$3/2$ & $3/2$,~$1$ &
$4/3$,~$3/2$ & $2$,~$1/2$ & $\infty$,~$0$ & $1/2$ & $0$, $1/2$\\

$~~$ Stochastic & $10/7$,~$7/4$ & $7/4$,~$1$ & $3/2$,~$3/2$ &
$5/2$,~$1/2$ & $4$,~$1/4$ & $3/4$ & $0$, $1/2$\\
\\

Non-Abelian\\
$~~$
Stochastic &
$\frac{2(3-\alpha)}{4-\alpha}$,~$2-\frac{\alpha}{2}$ &
$2-\frac{\alpha}{2}$,~$1$ &
$\frac{4+\alpha}{2(1+\alpha)}$,~$1+\alpha$ &
$\frac{2+\alpha}{2\alpha}$,~$\alpha$ &
$\frac{4(1-\alpha)}{2-3\alpha}$,~$1-\frac{3\alpha}{2}$ &
$1-\frac{\alpha}{2}$ & $\alpha$\\
\\
$~~~$ - numerics & $1.43(1)$,~$1.77(2)$ & $1.78(1)$,~$1.00(1)$ &
$1.53(1)$,~$1.46(2)$ & $2.74(2)$,~$0.44(1)$ &
$3.18(2)$,~$0.31(3)$ & $0.77(1)$ & $0.45(3)$  \\
\\

$~~$ Deterministic & $3/2$,~$2$ & $2$,~$1$ &
$\frac{2+\alpha}{1+\alpha}$,~$1+\alpha$ &
$\frac{1+\alpha}{\alpha}$,~$\alpha$ &
$\frac{2-\alpha}{1-\alpha}$,~$1-\alpha$ & $1$ & $\alpha$\\
\\
$~~~$ - aND & $1.49(1)$,~$1.97(3)$ & $1.94(1)$,~$1.00(1)$ &
$1.52(1)$,~$1.88(2)$ & $2.10(1)$,~$0.87(1)$ &
$6.64(3)$,~$0.07(1)$ & $0.96(1)$ & $0.86(3)$ \\

$~~~$ - bND & $1.43(1)$,~$1.82(4)$ & $1.79(1)$,~$1.00(1)$ &
$1.48(1)$,~$1.76(7)$ & $2.10(1)$,~$0.76(6)$ &
$5.90(9)$,~$0.06(1)$ & $0.81(2)$ & $0.69(5)$ \\

$~~~$ - cND & $1.52(3)$,~$1.99(4)$ & $2.04(3)$,~$1.00(1)$ &
$1.51(1)$,~$1.95(3)$ & $2.03(1)$,~$0.86(2)$ &
$9.26(7)$,~$0.06(4)$ & $0.99(2)$ & $0.91(11)$ \\

\hline\hline
\end{tabular}
\end{table*}

From $D_h=1-\alpha$ in the ND, we find that the avalanche
exponents for mass and duration have the mean-field (MF) values,
independent of $\alpha$, i.e. $\tau_s=3/2$ and $\tau_t=2$, whereas
other exponents depend on $\alpha$. We point out that one should
consider other avalanche exponents as well as those of mass and
duration in order to discuss the universality class of the ND. If
$\alpha=D_w=1$ is assumed from the linear behavior of avalanche
boundaries (scars) as shown in Fig.~\ref{fig:metastable}, all
avalanche exponents turn to the MF values, except for the case of
width~\cite{MFexceptw}. This may also correspond to the MF
behavior of coagulation-diffusion model, where the particle
density decays as $t^{-1}$ in $d\ge d_u=2$~\cite{Hinrichsen2000}.
This peculiar `MF' behavior of three ND versions appeared in the
low dimensional system can be understood by considering the shape
of $N(t)$ with its width and height. We now focus on how avalanche
boundaries behave linearly, which implies $D_w=1$. The ND toppling
rules we considered suppress the fluctuations of height profile of
the flowing avalanche more than the stochastic one does, which
leads to spread grains wider and make the avalanche boundaries
grow faster, almost ballistically. This positive feedback enables
$D_w=1$ to be larger than $1/2$ for all other DSMs. Moreover, we
like to note that the resultant $D_h=0$ indicates the MF behavior
for the non-Abelian case, whereas $D_h=0$ for any dimension in the
AD.

We performed extensive numerical simulations for all DSMs to
confirm our conjecture about the avalanche exponents in terms of
the scar exponent, $\alpha_{\rm sc}=\alpha$, up to $T=2^{13}$ and
$L=T/2$ ($T=2^{15}$ or $L=T$ in some cases). We measure the
avalanche exponent set $\{\tau_x,D_x\}$ of all $x$, $\alpha$, and
$\alpha_{\rm sc}$ using the moment analysis and the conventional
successive slope techniques of avalanche distributions, for about
$10^9$ avalanches at the steady state after the transient period.
The spatially correlated scars are observed in all DSMs with the
nonzero values of $\alpha_{\rm sc}$, while the spatial
correlations of metastable patterns are only observed in
non-Abelian DSMs with the nonzero values of $\alpha$. To validate
the stability of the scar exponent, we plotted the effective scar
exponent, $[\alpha_{\rm sc}]_{\rm eff}$, as a function of $1/t$
for all DSMs. As shown in Fig. 3(a), there seems to be two
asymptotic values, $1/2$ and $1$ in the large $t$ limit, with
quite long/unusaul initial transient behaviors and finite-size
corrections. For the non-Abelian case, we also checked the
possibility of logarithmic corrections to scaling in $\rho_{\rm
sc}$ and $\rho$, both of which behave qualitatively the same.
Thus, we only show $\rho$ in Fig. 3(b) as $\rho(t)\sim
t^{-\alpha}(\ln t)^{\phi}$, where the existence of linear parts in
curves represent logarithmic corrections.

We finally discuss the relevance of Abelian symmetry in DSMs.
Based on our results, Abelian symmetry turns out to be irrelevant
to the stochastic version only when $\alpha=1/2$ in the NS. The
breaking of Abelian symmetry in the deterministic version yields
the MF behavior of avalanche dynamics even in a (1+1)-dimensional
setup. In all NDs, we also confirm that the values of $D_h$ are
quite close to $0$, which can be the sign of the MF behavior as we
argued. Furthermore, it turns out the NDs do not show any
criticality in a (0+1)-dimensional setup. All numerical results
are listed in Table~\ref{table-exponent} with our conjecture. Only
the results of the bND seem to be inconsistent with the values we
conjectured. Such discrepancy may be attributed to the relevant
effect of bias in toppling rule or relatively large logarithmic
corrections. The validity of our conjecture for other possible
values of $\alpha$ or $\alpha_{\rm {sc}}$ is under investigation,
with the role of toppling bias in DSMs~\cite{ongoing,private}.

In summary, we have explored the role of Abelian symmetry and
stochasticity in directed sandpiles, and conjectured the new
scaling relations for critical avalanche dynamics entangled with
the underlying structure of metastable patterns and scars. Our
conjecture provides clear guidelines on discussing the
universality class in directed sandpile models. Moreover, our
results provide essential information on analyzing the
self-organized criticality in real systems as well as answering
how ubiquitous long-range spatial correlations in nature can be
developed and affect real avalanche dynamics.

This work was supported by the BK21 project and Acceleration
Research (CNRC)of MOST/KOSEF. M.H. would like to acknowledge
fruitful discussions with V.M. Uritsky and M. Paczuski, and the
kind hospitality of Complexity Science Group at the University of
Calgary, where the main idea of this work was initiated.




\begin{thebibliography}{00}
\bibitem{MFexceptw}
In general, on a ($d+1$)-dimensional lattice, $a\sim tw^d$
yielding $D_a=1+dD_w$. In case of $d\geq d_u$, where $d_u=2$ is
the upper critical dimension of DSMs, $dD_w=1$. This is consistent
with our `MF' results, i.e. $D_w=1$ in $d=1$.
\end{thebibliography}

\end{document}